\documentclass[prc, showpacs,amsmath,amssymb, floatfix]{revtex4}
\begin{document}
\hfill \fbox{\parbox[t]{1.12in}{LA-UR-05-7938}}\hspace*{0.35in}
\title{Scaling Factor Inconsistencies in Neutrinoless Double Beta Decay}
\author {S. Cowell} 
\affiliation{Theoretical Division, Los Alamos National Laboratory, 
Los Alamos, New Mexico, 87545, USA}

\date{\today}
\begin{abstract}
The modern theory of neutrinoless double beta decay includes a scaling factor
that has often been treated inconsistently in the literature.  The nuclear
contribution to the decay half life can be suppressed by $15$-$20$\% when
scaling factors are mismatched.  Correspondingly, $\langle m_{\nu} \rangle$ is
overestimated.
\end{abstract}

\maketitle

\section{Introduction}

In recent years, experimental evidence for neutrino masses and mixing have lead
to a concerted effort to refine the methods used in calculating reaction rates
for double-beta decay.   One of the largest uncertainties in these calculations
is the determination of the nuclear matrix elements (NME).   Together with
kinematic factors and experimental bounds on the decay half life, an average
neutrino mass can be extracted.  An accurate determination of the NME is
crucial and improvements in the QRPA and shell model techniques used to
calculate them continue to be explored.  

In the modern theory of neutrinoless double beta decay ($\beta \beta (0\nu)$)
definitions of the NME include a scaling factor introduced such that the NME is
dimensionless.  When using these scaled NME to determine the $\beta \beta
(0\nu)$ decay rate, the scaling factor must be compensated for elsewhere. 
However, the scaling factor has not always been treated consistently in the
literature.   In this article, we detail how the scaling factor is introduced
into the theory and illustrate the $15$-$20$\% suppression of the nuclear
contribution when mismatched scaling factors are used.

\section{Neutrinoless Double Beta Decay}
In the simplest form of the weak Hamiltonian
\begin{equation}
H_W = \frac{G}{\sqrt{2}}\left[ j_L^{\rho} J_{L \rho}^{\dagger} + h.c. \right]
\end{equation}
the half life of the $0^{+} \to 0^{+}$ $\beta
\beta(0\nu)$ is written as
\begin{equation}
[T^{(0\nu)}_{1/2}]^{-1} = |M^{(0\nu)}|^2 G_{01} \left(\frac{\langle m_{\nu}
\rangle}{m_e}\right)^2~.
\label{half}
\end{equation}
where $M^{(0\nu)}$ is the NME, $G_{01}$ is the
so-called phase space or kinematic factor and $m_{\nu}$ and $m_e$ are the
neutrino and electron masses respectively.  The NME is given by
\begin{eqnarray}
M^{(0\nu)} &=& M^{(0\nu)}_{GT}(1 - \chi_F) \\
M^{(0\nu)}_{GT} &=& \sum_a \langle 0^{+}_f || \sum_n \tau_n^{+} || N_a\rangle
\langle N_a || \sum_m \tau_m^{+} ||0^{+}_i \rangle (\sigma_n \cdot \sigma_m )
\frac{1}{2}(H_2-H_1) \\
\chi_F &=& \sum_a \langle 0^{+}_f || \sum_n \tau_n^{+} || N_a\rangle
\langle N_a || \sum_m \tau_m^{+} ||0^{+}_i \rangle 
\frac{1}{2}(H_2-H_1) \\.
\end{eqnarray}
The sum is taken over intermediate states $N_a$.  A detailed explanation of
the NME can be found in \cite{doi85}.  We will use notation adapted from 
\cite{doi85} throughout this article.

The $\nu$ potential induced by the virtual $\nu$ exchange is given by
\begin{eqnarray}
H_k(r, E_a) &\equiv& \frac{1}{2 \pi^2} \int \frac{d {\bf
q}}{\omega}\frac{1}{\omega + A_k} e^{i {\bf q}\cdot {\bf r}}~,
\label{pot}\\
A_{1(2)} &=& E_a - \frac{1}{2} (E_i + E_f)\pm \frac{1}{2}(\epsilon_1 -
\epsilon_2)~.
\end{eqnarray}
The initial, intermediate and final state energies are denoted by $E_i$, $E_a$
and $E_f$ respectively and $\epsilon_i$ denotes the energy of the $i$th
electron.

The kinematic factor $G_{01}$ in Eq. (\ref{half}) is defined as
\begin{eqnarray}
G_{01} &=& \frac{a_{0\nu}}{m_e^2 ~ \mathrm{ln}2} \int d\Omega_{0\nu}
~F_0(Z,\epsilon_1)~ F_0(Z,\epsilon_2)~,
\label{kf}
\end{eqnarray}
where
\begin{eqnarray}
a_{0\nu} &=& \frac{(G g_A)^4 m_e^9}{64 \pi^5}\\
d\Omega_{0\nu} &=& m_e^{-5}~ p_1~ p_2 ~\epsilon_1~ \epsilon_2~ \delta(\epsilon_1 +
\epsilon_2 + E_f - E_i) ~d\epsilon_1 ~d\epsilon_2 ~d(\hat{{\bf p}}_1\cdot\hat{{\bf
p}}_2)~.
\end{eqnarray}
We have assumed $S=0$ electron wave functions with no $r$ dependence. The
Fermi functions, $F_0(Z,\epsilon)$, depend upon the charge of the daughter
nucleus, $Z$, and the energy of the $i$th electron, $\epsilon_i$.

In the early $\beta \beta(0\nu)$  calculations, the NME and kinematic factor
were defined as above.  The NME were given in units of fm$^{-1}$ and the
kinematic factors in units of yr$^{-1}$ fm$^2$ (for example,
\cite{tomoda86}).  In the mid-eighties, a scaling factor was introduced into
the $\beta \beta(0\nu)$ theory.   The $\nu$ potential, Eq. (\ref{pot}), is
scaled by a factor of $R=r_0 A^{1/3}$ such that the NME are dimensionless:  
\begin{eqnarray}
h_{+} &=& \frac{R}{2}(H_2+H_1) \label{nupot}\\
\tilde{M}^{0\nu} &=& \tilde{M}^{(0\nu)}_{GT}(1-\tilde{\chi}_F)\\
\tilde{M}^{(0\nu)}_{GT} &=& \sum_a \langle 0^{+}_f || \sum_n \tau_n^{+} || N_a\rangle
\langle N_a || \sum_m \tau_m^{+} ||0^{+}_i \rangle (\sigma_n \cdot \sigma_m )
h_{+}\\
\tilde{\chi}_F &=& \sum_a \langle 0^{+}_f || \sum_n \tau_n^{+} || N_a\rangle
\langle N_a || \sum_m \tau_m^{+} ||0^{+}_i \rangle 
h_{+}\\.
\end{eqnarray}

The scaling factor, $R$, in $h_{+}$ is compensated for by introducing $1/R^2$
into the definition of the kinematic factor, Eq. (\ref{kf}):  
\begin{equation}
G^S_{01}(R) = \frac{a_{0\nu}}{(m_e R)^2 ~\mathrm{ln}2} \int d\Omega_{0\nu}
~F_0(Z,\epsilon_1)~ F_0(Z,\epsilon_2)
\label{kf2}
\end{equation}

The $G^S_{01}(R)$ have been calculated by many authors; for example,
\cite{doi85} using $R = 1.2 A^{1/3}$ and \cite{pantis96} using $R = 1.1
A^{1/3}$. Though the underlying physics of the kinematic factor is unchanged,
the published values of $G_{01}^S(R)$ are significantly different due to
different choices of $R$.  Provided that the NME and kinematic factor are
calculated using the same scaling factor, these differences are irrelevant. 
However, the $R$ used in calculating the NME have not always been carried
consistently to the kinematic factor.

In recent years, numerous calculations of NME have been performed using
$h_{+}$.  As techniques develop, it is customary to compare with 
previously published values.  These comparisons are often made by defining
$C_{00}$:
\begin{eqnarray}
[T_{1/2}^{(0\nu)}]^{-1} &=& C_{00} \left(\frac{\langle m_{\nu}
\rangle}{m_e}\right)^2 \\
C_{00} &=& |\tilde{M}^{(0\nu)}|^2 G_{01}^S(R)
\label{coo}
\end{eqnarray}
If scaling is treated consistently, $C_{00}$ is independent of the scaling
factor.  However, in citing previous calculations of the NME, the scaling
factor has not always been accounted for properly. Often, 
$G_{01}^S(R=1.2 A^{1/3})$ is used to calculate $C_{00}$ regardless of the
scaling factor used in determining the NME.  If one combines NME calculated
with $R=1.1 A^{1/3}$ fm with $G_{01}^S(R=1.2 A^{1/3})$,  $C_{00}$ is suppressed
by $(1.1/1.2)^2 \sim 20$\%.  Correspondingly, this leads to an overestimation
of $\langle m_{\nu} \rangle$.

Table \ref{table1} includes several $C_{00}$ predictions for $^{76}$Ge $\beta
\beta (0\nu)$.  This table was originally published in a recent review article
\cite{elliott04} and adapted from \cite{civitarese03}.   The mismatch of
scaling factors has occurred several times in the literature.  We include
revisions to Table 2 of \cite{elliott04} because it is one of the more thorough
listings of NME calculations.  

In Table \ref{table1} we include both the previously published $C_{00}$ and the
revised values obtained using the correctly scaled  $G_{01}^S(R)$.  For
clarity, the value of $r_0$ used in the original publication is included.  In
some instances $r_0$ was not clearly stated; the $r_0$ value was extracted from
$T_{1/2}$, $\langle m_{\nu} \rangle$ or stated $G_{01}^S(R)$ values given in
the original paper assuming that the scaling was originally treated
consistently.  

Using the simple $S=0$ electron wave functions with no $r$ dependence, we
obtain $G_{01}^S(R=1.2 A^{1/3}) = 6.46 \cdot 10^{-15}$ yr$^{-1}$ and
$G_{01}^S(R=1.1 A^{1/3}) = 7.78 \cdot 10^{-15}$ yr$^{-1}$.  These values are
within a few percent of those published previously by \cite{doi85} and
\cite{pantis96}. Using the appropriate kinematic factor, most $C_{00}$
determined using NME with $r_0=1.1$ are changed by $\approx 20$\%. Assuming a
half life of $4 \cdot10^{27}$ yr the spread of predicted $\langle m_{\nu}
\rangle$ values is unchanged, 0.022-0.068.  However, several predicted neutrino
masses are reduced.

\begin{table}[htb]
\begin{center}
\begin{tabular}{l|l|c|c|l|l}
$C_{00}^{\mathrm{revised}}$ & $C_{00}^{\mathrm{old}}$ & 
$\langle m_{\nu} \rangle $ & Method & $r_0$ & Reference\\
$~\mathrm{x}~10^{-14} (\mathrm{yr}^{-1})$ & $~\mathrm{x}~10^{-14} (\mathrm{yr}^{-1})$ & 
(eV)& & (fm) & \\
\hline
$11.2 $ & 11.2 & 0.024 & QRPA & 1.2$^{2}$& \cite{muto89,staudt90} \\
$6.97 $ & 6.97 & 0.031 & QRPA & NS & \cite{suhonen92} \\
$7.51 $ & 7.51 & 0.029 & Number-projected QRPA & NS & \cite{suhonen92} \\
$7.19 $ & 7.33 & 0.030 & QRPA & 1.1 & \cite{pantis96} \\
$12.1 $ & 12.0 & 0.023 & QRPA & NS & \cite{tomoda91} \\
$13.3 $ & 13.3 & 0.022 & QRPA & NS & \cite{aunola98} \\
$8.34 $ & 8.27 & 0.028 & QRPA & 1.2 & \cite{barbero99} \\
$1.89-12.8 $ & 1.85-12.5 & 0.059-0.023 & QRPA & 1.2$^1$ & \cite{stoica01} \\
$5.02-5.93 $ & 1.8-2.2 & 0.036-0.033 & QRPA & 1.1$^1$ & \cite{bobyk01} \\
$8.61 $ & 8.36 & 0.028 & QRPA & 1.2$ ^1$ & \cite{civitarese03} \\
$1.40 $ & 1.42 & 0.068 & QRPA with np pairing & 1.1 & \cite{pantis96} \\
$5.59 $ & 4.53 & 0.034 & QRPA with forbidden & 1.1$^1$ & \cite{rodin03} \\
$10.1 $ & 10.3 & 0.025 & RQRPA & 1.1 & \cite{simkovic99} \\
$6.10 $ & 6.19 & 0.033 & RQRPA with forbidden & 1.1 & \cite{simkovic99} \\
$6.77-7.72 $ & 5.5-6.3 & 0.031-0.029 & RQRPA & 1.1$^1$ & \cite{bobyk01} \\
$2.26-9.04 $ & 2.21-8.83 & 0.054-0.027 & RQRPA & 1.2$^1$ & \cite{stoica01} \\
$4.48 $ & 3.63 & 0.038 & RQRPA with forbidden & 1.1$^1$ & \cite{rodin03} \\
$2.71 $ & 2.75 & 0.049 & Full RQRPA &  1.1 & \cite{simkovic97} \\
$3.72-8.75 $ & 3.36-8.54 & 0.042-0.027 & Full RQRPA & 1.2$^1$  & \cite{stoica01} \\
$6.66-9.43 $ & 6.50-9.21 & 0.031-0.026 & Second QRPA & 1.2$^1$ & \cite{stoica01} \\
$0.34-0.40 $ & 0.27-0.32 & 0.139-0.128& Self-consistent QRPA* & 1.1$^1$ & \cite{bobyk01} \\
$28.8 $ & 28.8 & 0.015 & VAMPIR* & NS & \cite{tomoda86} \\
$15.6 $ & 15.8 & 0.020 & Shell model truncation* & NS & \cite{haxton84} \\
$7.03-16.2 $ & 6.87-15.7 & 0.030-0.020 & Shell model truncation* & 1.2$^1$   & \cite{engel89} \\
$1.94 $ & 1.90 & 0.058 & Large-scale shell model & 1.2$^1$ & \cite{caurier96}
\end{tabular}
\end{center}
\caption{A comparison of the nuclear contributions to $T_{1/2}$ , $C_{00}$,
and the resulting $\langle m_{\nu} \rangle$ values assuming $T_{1/2}=4.0 \cdot
10^{27}$ yr.  Included are the values of $r_0$ used in each NME calculation. 
The $r_0$ has been explicitly stated in the original publication unless
otherwise indicated as:  NS (no scaling/$C_{00}$ published in original paper);
 $^1$ ($r_0$
inferred from published values of $\langle m_{\nu}\rangle$, $T_{1/2}$ or
$G_{01}$) and $^2$ (Published $C_{00}$; $r_0$
inferred from published values of $\langle m_{\nu}\rangle$, $T_{1/2}$ or
$G_{01}$)\label{table1}}
\end{table}

It is important to point out $G_{01}$, defined without the scaling factor, does
depend on $R$ through the Fermi functions of the electron wave functions.  The
appropriate choice of $R$ depends upon the nucleus being considered and should
be chosen such that experimental values of the mean square radius $\langle
r^2\rangle$ are reproduced. For a uniform charge distribution, $R^2 = 5/3
\langle r^2\rangle$.  In Table \ref{table2} we give the \emph{unscaled}
$G_{01}$ calculated using $R$ values fit to experimental root-mean-square radii
when possible \cite{devrie}.  Comparing $G_{01}$ for $^{150}$Nd, it is clear
that the unscaled $G_{01}$ are not very sensitive to the choice of $R$.  The
significant differences obtained by \cite{doi85} and \cite{pantis96} for
$G_{01}^S(R)$ are predominately due to the scaling factor.

\begin{table}[]
\begin{center}
\begin{tabular}{c|ccc}
Nucleus &	 $\qquad \langle r^2 \rangle^{1/2}\qquad $ & $ \qquad r_0\qquad$ 
& $\qquad G_{01}^{0\nu}\qquad$ \\
 &	 fm & fm & fm$^2$ yr$^{-1}$\\
\hline
$^{ 48}$Ca	& 3.470 & 	1.23 & $1.236x 10^{-12 }$\\
		& 3.451 &  	1.23 & $1.236x 10^{-12 }$\\
$^{ 76}$Ge 	& 4.081 & 	1.24 & $1.663x 10^{-13 }$\\
$^{ 82}$Se 	& no data & 		1.20 & $7.784x 10^{-13 }$\\
$^{ 96}$Zr 	& 4.396 & 	1.24 & $1.792x 10^{-12 }$\\
$^{100}$Mo 	& 4.430 & 	1.23 & $1.436x 10^{-12 }$\\
$^{116}$Cd 	& 4.639 & 	1.23 & $1.720x 10^{-12 }$\\
$^{128}$Te 	& no data & 		1.20 & $6.609x 10^{-14 }$\\
$^{130}$Te 	& no data & 		1.20 & $1.661x 10^{-12 }$\\
$^{136}$Xe 	& no data & 		1.20 & $1.825x 10^{-12 }$\\
$^{150}$Nd 	& 5.048 & 	1.23 & $8.719x 10^{-12 }$\\
		& 5.015 &  	1.22 & $8.750x 10^{-12 }$\\
		& 4.948 &  	1.20 & $8.813x 10^{-12}$
\end{tabular}
\end{center}
\caption{$0^{+} \to 0^{+}$  $\beta \beta(0\nu)$ kinematic factor $G_{01}$
calculated at the specified values of $r_0$  obtained by fitting
the experimental $\langle r^2 \rangle^{1/2}$
when available \cite{devrie}. \label{table2}}
\end{table}

To avoid confusion in the future, we strongly encourage that further
calculations of NME be published with either no scaling factor included, or a
clear indication of what choice the authors have made for $r_0$.  

\begin{acknowledgments} 
The author would like to thank Steve Elliott for suggesting the need to
resolve this issue as well as A.C. Hayes, J.L. Friar and J. Carlson
for several useful discussions.
\end{acknowledgments}

\end{document}